\documentclass[aps,prc,twocolumn,groupaddress,showkeys]{revtex4-1}
\usepackage{graphicx} 
\usepackage{dcolumn} 
\usepackage{bm}
\usepackage{color}
\usepackage{hyperref}

\begin{document}

\title{Probing nuclear bubble configuration by proton induced reaction}
\author{Xiao-Hua Fan$^{1,2,3}$}
\author{Gao-Chan Yong$^{1,3}$}
\email[]{yonggaochan@impcas.ac.cn}
\author{Wei Zuo$^{1,3}$}
\affiliation{$^1$Institute of Modern Physics, Chinese Academy of Sciences, Lanzhou 730000, China\\
$^2$School of Physical Science and Technology, Lanzhou University, Lanzhou 730000, China\\
$^3$School of Nuclear Science and Technology, University of Chinese Academy of Sciences, Beijing 100049, China}

\begin{abstract}

In the framework of the isospin-dependent Boltzmann-Uehling-Uhlenbeck transport model, nuclear bubble configuration in the hypothetical $^{48}$Si nucleus is studied by proton induced central reaction at an incident beam energy of 0.8 GeV/nucleon. It is found that along the beam direction more energetic protons are emitted with bubble configuration in the target. In the forward angles, compared with the case without bubble configuration, less scattered energetic protons are emitted with bubble configuration in the target. We thus provide a new way to probe the bubble configuration in nuclei.


\end{abstract}

\maketitle


The nucleon spatial density distribution is one of the most fundamental  characters  of atomic nucleus structure. On the account of the nuclear force's saturation  properties, it's generally considered that a nucleus is commonly compact and the radial locations of neutrons and protons in finite nuclei are in the form of Fermi distributions. However, the density profile could have another atypical forms, such as halo and central depletion, due to the complexity of the nuclear force and quantum many-body system \cite{halo1,halo2,bub1,bub2,bub3}.  The central depletion is vividly called bubble or hollow, which is dependent on the depression range in nuclear density.   It has been theoretically discussed that the existence of a bubble structure inside a superheavy  nucleus can decrease the total energy of the nucleus \cite{bub1,bub2,bub3}. In fact, the bubble-like structure was first suggested in 1946 \cite{bub4}, in which the nucleus was assumed to be a thin spherical shell to explain a series of nuclei with equally spaced energy levels. Two decades afterwards, the spherical bubble nuclei were studied based on liquid drop model (LDM) in 1967 \cite{1967}. And later on, the bubble occurrences were extensively explained by utilizing several approaches, such as LDM plus shell correction energy \cite{wong}, the Hartree-Fock method \cite{Campi}, the Thomas-Fermi model \cite{Saunier} as well as transport simulations \cite{yong}. It is interesting to note that the existence of the bubble structure is not limited to a certain region but a quite extensive region of the nuclear chart \cite{light1,light2,dong1,suph1,bub1,bub2,bub3}. And transport simulations show that general excited heavy nuclei may also have central depleted density  \cite{yong}.

It is exciting to see that the first experimental evidence that points to a depletion of the central density of protons in the short-lived nucleus $^{34}$Si was provided in 2016 \cite{nature2016}. Also in 2016, Najman \emph{et al.} have recently found some signatures of exotic nuclear configurations such as toroid-shaped objects in $^{197}$Au + $^{197}$Au reaction at 23 MeV/nucleon \cite{Najman}.

Using the relativistic Hartree-Fock Lagrangian PKA1, $^{48}$Si is predicted to be the first candidate of dual semi-bubble nucleus with both neutron and proton bubble-like shapes \cite{long2019}. Since the existence or not of the dual semi-bubble nucleus $^{48}$Si can help us to deeply understand the nature of nuclear force, it is very necessary to find an effective way to confirm the dual semi-bubble nucleus $^{48}$Si. Of course, our present study based on the bubble nucleus $^{48}$Si in proton induced reaction is also, qualitatively or quantitatively, applicable to some other potential bubble or semi-bubble nuclei, such as $^{34}$Si, etc.

There are many sensitive probes in heavy-ion collisions to probe nuclear structure and property of both finite nuclei and nuclear matter. Those probes are also considered as an expectation to investigate the bubble configuration. Actually, the visualized probes which can be useful to recognize the novel density profiles have been discussed in recent yeas \cite{Bauer,Gross,xu,li,Cherevko,yong}. In the framework of the isospin-dependent Boltzmann-Uehling-Uhlenbeck (IBUU) transport model, the bubble configuration in $^{48}$Si is studied in the central p+$^{48}$Si reaction at 0.8 GeV/nucleon. It is found that in proton induced reaction there are evident effects of bubble configuration on the energetic proton emission in forward angles. The energetic proton emission as a function of scattering angle thus can be a probe of nuclear bubble configuration in the target in proton induced reaction.


The used isospin-dependent Boltzmann-Uehling-Uhlenbeck transport model has evident isospin-dependent nuclear initialization, isospin-dependent nucleon-nucleon cross sections, isospin-dependent single nucleon potential and isospin-dependent Pauli blockings. The used single nucleon potential includes a Skyrme-type parametrization  isoscalar term and an exponential isovector term \cite{YONG2}. This model has been successfully used to study the bubble configuration of nucleon matter recently \cite{yong,YONG2}.
The relativistic Hartree-Fock-Bogoliubov (RHFB) with the PKA1 Lagrangian is one of the most advanced relativistic approaches which is apt at delineating the properties of very neutron-rich nuclei \cite{long2019,long2}. The dual semi-bubble nucleus $^{48}$Si is one possible candidates for double-bubble (both proton and neutron bubble structures) nucleus. For comparison,
initializations of the nucleon density distribution in the used transport model adopted two different methods, one uses the general Skrymer-Hartree-Fock (SHF) with
Skyrme M* force parameters \cite{skm} and the other uses the PKA1 Lagrangian \cite{long2019,long2}.

\begin{figure}[tbh]
\includegraphics[width=9 cm]{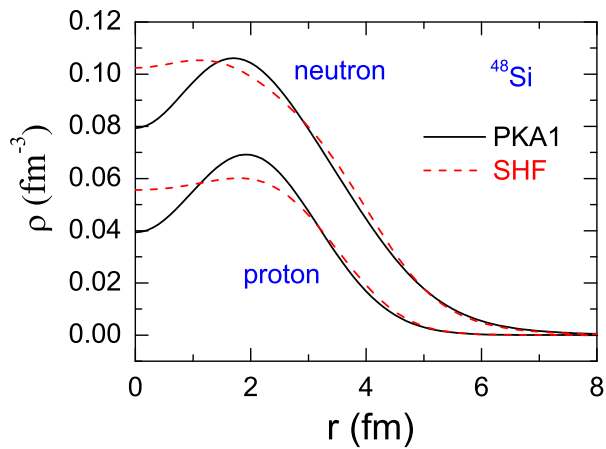}
\caption{\label{fig1} (Color online) The density distributions of neutrons and protons in the
ground state for $^{48}$Si, calculated by Skrymer-Hartree-Fock (SHF) and the PKA1 Lagrangian of the relativistic Hartree-Fock-Bogoliubov(RHFB).}
\end{figure}
Fig. \ref{fig1} shows the density distributions of neutrons and protons  as a function of radial co-ordinates in the ground state for $^{48}$Si obtained by the two approaches mentioned above. The neutron density is much higher than the proton's for both of the two cases on account of the  great neutron-richness of $^{48}$Si. It's apparently seen that with PKA1 there is a true bubble configuration with a depletion density at the center for both neutrons and protons, while there is a quite compact center for nucleons with the SHF.

\begin{figure}[tbh]
\includegraphics[width=9 cm]{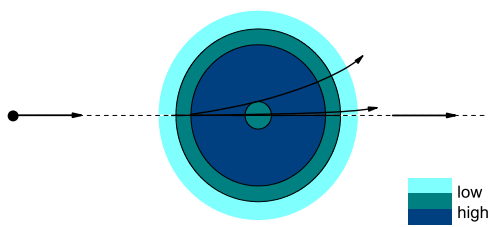}
\caption{\label{fig2} (Color online) The sketch of the reaction with proton as projectile and $^{48}$Si as target. The depth of color represents the denseness of baryonic density, namely the density increases from low to high as the color changes from light blue to dark blue.}
\end{figure}
The sketch of the reaction with proton as projectile and the bubble nucleus $^{48}$Si with density distribution given by PKA1 as target is shown in Fig. \ref{fig2}.
It displays the scattering situations in the X-Z profile (Z-axis is the beam direction and X-axis is perpendicular to the Z-axis). The proton is directly projected to the center of $^{48}$Si along the Z axis.  The depth of color represents the denseness
of baryonic density, namely the density of $^{48}$Si becomes larger as the color changes from light blue to dark blue. Because of the bubble structure inside $^{48}$Si, the reaction should be different from that without bubble in the target. More detailed information about the reaction needs some resultful simulations.

\begin{figure}[h!]
\includegraphics[width=9 cm]{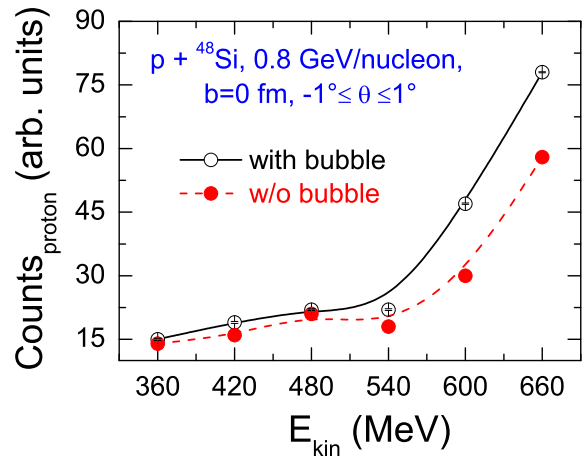}
\caption{\label{fig3} (Color online) Counts of emitted protons as a function of kinetic energy detected in the beam direction in the central p+$^{48}$Si reaction at 0.8 GeV/nucleon with (RHFB+PKA1) and without (SHF) bubble configuration in the target.}
\end{figure}
Besides the rough sketch explaining the scattering process physically to some extent, the central p+$^{48}$Si reaction at a beam energy of 0.8 GeV/nucleon and an impact parameter of 0 fm is simulated based on the framework of the IBUU transport model with density distributions initialized by the SHF (without bubble configuration in the target) and the PKA1 (with bubble configuration in the target). The central collision generally corresponds to the high multiplicity events experimentally. To determine the impact parameter more precisely, the neural network method may be used to select the most central experimental reactions by using the measured values of some observables (such as the multiplicity of charged particles, the transverse and longitudinal momentum distributions of outgoing particles) as the input variables \cite{b0fm}. Hitting the target $^{48}$Si by energetic proton, there should be plenty of emitted protons. The count of emitted protons as a function of kinetic energy detected in the beam direction ($-1^\circ\leq\theta\leq1^\circ$, $\theta$ is the angle between emitted direction and the Z axis) in the central p+$^{48}$Si reaction with and without the bubble configuration is plotted in Fig. \ref{fig3}. To stabilize the result, for each case a million events are accumulated. It is seen that the number of emitting protons with the bubble configuration in the target is evidently larger than that without the bubble configuration in the target, especially for energetic emitting protons. This is because energetic protons are easier to go through the target with the bubble configuration in the target compared with the case without the bubble configuration. From Fig. \ref{fig3}, it is also seen that the effects of the bubble configuration in the target on the energetic emitting protons ($\text{E}_{\text{kin}}\geq$ 600 MeV) can reach 30--40\%. Such large effect of the bubble configuration in the target can be  detected experimentally.

Although the dual semi-bubble nucleus $^{48}$Si is currently hard to produce \cite{long2019}, a potential bubble nucleus with similar central depletion density can be an alternative. If they roughly have the same depletion of the central density, the effect of the bubble configuration in the target on the energetic emitting protons should be also similar.

\begin{figure}[tbh]
\includegraphics[width=9 cm]{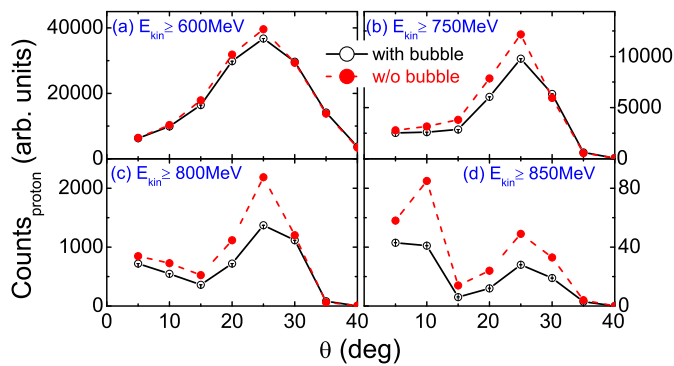}
\caption{\label{fig4} (Color online) Counts of emitted protons as a function of scattering angle (relative to the beam direction) in the central p+$^{48}$Si reaction at 0.8 GeV/nucleon with and without bubble configuration in the target. Different panels denote different kinetic energy cuts.}
\end{figure}
Besides counting energetic emitting protons in the beam direction in proton induced reaction, it is also interesting to see the counts of emitted protons with different scattering angles. Fig. \ref{fig4} shows the number of emitted energetic protons as a function of scattering angle (relative to the beam direction) in the central p+$^{48}$Si reaction at 0.8 GeV/nucleon with and without bubble configuration in the target. Each panel has a specific kinetic energy cut.
From Fig. \ref{fig4}, it is seen that, in forward angles (especially $10^{\circ}\leq\theta\leq25^{\circ}$) the number of scattered energetic protons without bubble configuration in the target is larger than the case with the bubble configuration in the target. However, the bubble effect disappears with the increase of the scattering angle. This is because in the forward angle zone, more energetic protons are scattered with compact central density in the target (i.e., without bubble configuration in the target). As the scattering angle increases, the effects of the bubble configuration in the target on the energetic scattered protons decrease due to the small proportion of the bubble area in the whole target. From Fig. \ref{fig4}, it is also seen that the effects of the bubble configuration in the target reach maximum around $\theta= 22.5^{\circ}$. Here the specific value of the scattering angle should be decided by the specific size of the bubble and the size of target nucleus.

For scattering energetic protons, the effects of the bubble configuration in the target in proton induced reaction are also decided by the kinetic energy of emitting protons. From panels (a)-(d) in Fig. \ref{fig4}, it is seen that the effects of the bubble configuration on the energetic scattered protons in forward angles reach 16--70\% with kinetic energy cuts from E$_{\text{kin}}>$ 600 MeV to E$_{\text{kin}}>$ 850 MeV. It shows that very energetic scattering protons in forward angles in proton induced reaction are more suitable to probe the bubble structure in the target. This is because the energetic scattered protons are less affected by the Fermi momentum of nucleons in nucleus thus show more effects of the bubble structure in colliding target.

The present analytical method of the proton induced reaction that is used to probe the bubble configuration in the hypothetical $^{48}$Si nucleus can of course be used to probe some other potential bubble or semi-bubble nuclei \cite{light1,grasso09,dug17}. And the proton induced reaction to probe the bubble configuration in nuclei can be considered as an alternative of  using the electron scattering experiments to probe nuclear bubble, such as at ELISe@FAIR \cite{ec17} or after an upgrade of the SCRIT facility at RIKEN \cite{ec18}.


The bubble configuration of atomic nuclei, which is characterized by a depletion of their central density, has been discussed for many decades. The discovery of bubbles in nuclei is an important issue for nuclear structure and nuclear many-body approach. Based on the isospin-dependent Boltzmann-Uehling-Uhlenbeck transport model, nuclear bubble configuration in the hypothetical $^{48}$Si nucleus is studied by proton induced central reaction at an incident beam energy of 0.8 GeV/nucleon. It is found that the emitted protons are much more in the beam direction when the target has a hollow structure than does not have it while in the forward angles (nonzero angles) energetic protons are less scattered when the target has a hollow structure than does not have it. Both effects are enhanced for higher energy proton emission. Angle distribution of the energetic proton emission in the proton induced central reaction at a beam energy of 0.8 GeV/nucleon thus can be a probe of the bubble configuration in target. The present study can also be used to probe some other potential bubble nuclei. Our present results act as a
strong motivation to probe some potential bubble nuclei in future proton induced nuclear reaction experiments.


The author XHF thanks Prof. W.H. Long for helpful discussions.
This work is supported in part by the National Natural Science
Foundation of China under Grant Nos. 11775275, 11435014 and the 973 Program of China (2007CB815004).


\begin{thebibliography}{99}

\bibitem{halo1}A. M. Poskanzer, S. W. Cosper, and Earl K. Hyde,  Phys. Rev. Lett. {\bf 17}, 1271 (1966).
\bibitem{halo2}A. S. Jensen, K. Riisager, D. V. Fedorov, and E. Garrido, Rev. Mod. Phys. {\bf 76}, 215 (2004).
\bibitem{bub1}J. Decharg\'e, J. F. Berger, K. Dietrich, and M. S. Weiss,  Phys. Lett. B {\bf 451}, 275 (1999).
\bibitem{bub2}W. Nazarewicz et al., Nucl. Phys. A {\bf 701}, 165c (2002).
\bibitem{bub3}J. Decharg\'e, J. F. Berger, M. Girod, and K. Dietrich,  Nucl. Phys. A {\bf 716}, 55 (2003).
\bibitem{bub4}H. A. Wilson, Phys. Rev. {\bf 69}, 538 (1946).
\bibitem{1967}P. J. Siemens and H. A. Bethe, Phys. Rev. Lett. {\bf 18}, 704 (1967).
\bibitem{wong}C. Y. Wong, Ann. Phys. (NY) {\bf 77}, 279 (1973).
\bibitem{Campi}X. Campi and D. W. L. Sprung, Phys. Lett. B {\bf 46}, 291 (1973).
\bibitem{Saunier}G. Saunier, B. Rouben, and J. Pearson, Phys. Lett. B {\bf 48}, 293 (1974).
\bibitem{yong}G.C. Yong, Phys. Rev. C {\bf 93}, 014602 (2016).
\bibitem{light1}E. Khan, M. Grasso, J. Margueron, and N. Van Giai, Nucl. Phys. A {\bf 800}, 37 (2008).
\bibitem{light2}H. Nakada, K. Sugiura, and J. Margueron, Phys. Rev. C {\bf 87}, 067305 (2013).
\bibitem{dong1}Y. Z. Wang, J. Z. Gu, X. Z. Zhang, and J. M. Dong, Phys. Rev. C {\bf 84}, 044333 (2011).
\bibitem{suph1}J. Decharge\'e, J.-F. Berger, K. Dietrich, and M. Weiss, Phys. Lett. B {\bf 451}, 275 (1999).
\bibitem{nature2016}A. Mutschler, A. Lemasson, O. Sorlin et al., Nature Physics {\bf 13}, 152 (2017).
\bibitem{Najman}R. Najman et al., Phys. Rev. C {\bf 92}, 064614 (2015).

\bibitem{long2019}J. J. Li, W. H. Long, J. Margueron, and N. V. Giai, Phys. Lett. B {\bf 788}, 192 (2019).

\bibitem{Bauer}W. Bauer, G. F. Bertsch, and H. Schulz, Phys. Rev. Lett. {\bf 69}, 1888 (1992).

\bibitem{Gross}D. H. E. Gross, B. A. Li, and A. R. DeAngelis, Ann. Phys.(NY) {\bf 504}, 467 (1992).
\bibitem{xu}H. M. Xu, J. B. Natowitz, C. A. Gagliardi, R. E. Tribble, C. Y. Wong, and W. G. Lynch, Phys. Rev. C {\bf 48}, 933 (1993).
\bibitem{li}B. A. Li, and D. H. E. Gross, Nucl. Phys. A {\bf 554}, 257 (1993).
\bibitem{Cherevko}K. Cherevko, L. Bulavin, J. Su, V. Sysoev, and F. S. Zhang, Phys. Rev. C {\bf 89}, 014618 (2014).
\bibitem{YONG2}G. C. Yong, Eur. Phys. J. A {\bf 52}, 118 (2016).
\bibitem{long2}J. J. Li, W. H. Long, J. L. Song, and Q. Zhao, Phys. Rev. C {\bf 93}, 054312 (2016).
\bibitem{skm}J. Friedrich, and P.G. Reinhard, Phys. Rev. C {\bf 33} 335 (1986).
\bibitem{b0fm}C. David, M. Freslier, and J. Aichelin, Phys. Rev. C {\bf 51} 1453 (1995).
\bibitem{grasso09}M. Grasso, L. Gaudefroy, and E. Khan et al., Phys. Rev. C {\bf 79}, 034318 (2009).
\bibitem{dug17}T. Duguet, V. Som\`{a}, S. Lecluse, C. Barbieri, and P. Navr\'{a}til, Phys. Rev. C {\bf 95}, 034319 (2017).

\bibitem{ec17}H. Simon, Nucl. Phys. A 787, {\bf 102} (2007).
\bibitem{ec18}T. Suda, M. Wakasugi, and T. Emoto et al., Phys. Rev. Lett. {\bf 102}, 102501 (2009).



\end{thebibliography}
\end{document}